\documentclass[prd,twocolumn,showpacs,nofootinbib,preprintnumbers]{revtex4}

\usepackage{amsmath}
\usepackage{amsfonts}
\usepackage{graphicx}
\usepackage{dcolumn}
\usepackage[T1]{fontenc}
\usepackage{hyperref}

\def\be{\begin{equation}}
\def\ee{\end{equation}}
\def\ba{\begin{eqnarray}}
\def\ea{\end{eqnarray}}
\def\bs{\begin{subequations}}
\def\es{\end{subequations}}
\newcommand{\vp}{\varphi}
\newcommand{\ve}{\varepsilon}
\newcommand{\dth}{\bar{\theta}}

\newcommand{\dy}{\bar{y}}
\newcommand{\da}{\bar{a}}
\newcommand{\de}{\bar{\epsilon}}
\newcommand{\dH}{\bar{H}}
\newcommand{\dalp}{\bar{\alpha}}

\begin{document}

\title{Regularized dualities in patch cosmology}
\author{Gianluca Calcagni}
\email{calcagni@fis.unipr.it}
\affiliation{Dipartimento di Fisica, Universit\`{a} di Parma\\}
\affiliation{INFN -- Gruppo collegato di Parma\\ Parco Area delle Scienze 7/A, I-43100 Parma, Italy}
\date{October 21, 2004}
\begin{abstract}
Following past investigations, we explore the symmetries of the Hamilton-Jacobi cosmological equations in the generalized patch formalism describing braneworld and tachyon scenarios. Dualities between different patches are established and regular dual solutions, either contracting or phantom-like, are constructed.
\end{abstract}
\preprint{gr-qc/0410xxx}
\preprint{UPRF-2004-26} 
\maketitle


Recently a lot of attention has been devoted to the symmetries of the cosmological dynamics. Transformations of the Einstein equations, encoded in the Friedmann relation coupled to the equations of motion for the matter content in the Universe, link standard inflationary cosmologies to other possible phases \cite{BGV,wan99,chi02,GKST,KST,ACJL,CL,DSS,BST,ACL,PZ,pia04,lid04}. These are either contracting periods, ideally embedded in some motivated high-energy pre-inflationary framework, or superaccelerating cosmologies, $\ddot{a}/a>H^2$, dominated by a matter component (called ``phantom'') with and effective equation of state $p<-\rho$. Such scenarios are of particular interest from both a theoretical and observational point of view, since the first one is intertwined with the big bang problem and the resolution of the initial singularity, while phantoms might explain modern data on the late-time evolution of the Universe. Conversely, bouncing events can leave their imprint on the large-scale perturbation spectra, while a phantom component can arise in a stringy or supersymmetric setup.

In four dimensions, these dualities (inflation-contraction, inflation-phantom and contraction-phantom) are exact, in the sense that not only the dynamics but also the cosmological spectrum of scalar perturbations is preserved. The issue has then been generalized to the braneworld context \cite{cal9,CLLM}. Since braneworld spectra are broken under duality, and mapped into a quantitatively different contracting-like or phantom-like spectra, these transformations are not symmetries in the strict meaning of the word.

In this Letter we shall use the results of \cite{cal9} and extend them to dualities between different patches, which we will call cross dualities. Then, by considering the general invariance of the patch Hamilton-Jacobi equations we will be able to construct dual solutions with canonical behaviour, that is not suffering sudden future singularities. The starting point is the Friedmann equation
\be \label{FRW}
H^2=\beta^2 \rho^q\,,
\ee
valid in some particular energy regime (a ``patch'') or time interval during the evolution of the braneworld. Here $H$ is the effective Hubble rate experienced by an observer on the brane, $\rho$ is the effective energy density related to pressure by the equation of state $p=w\rho$, $\beta$ is a constant and the exponent $q$ is equal to 1 in the pure four-dimensional general-relativistic regime, $q=2$ in the high-energy limit of the Randall-Sundrum (RS) braneworld and $q=2/3$ in the high-energy limit of the Gauss-Bonnet (GB) scenario.

During inflation, the cosmological evolution is triggered by a scalar field $\vp$ with potential $V(\vp)$ and equations of motion
\ba \label{hjphi}
V(\phi) &=& \left(1-\frac{\epsilon}{3q}\right)|H|^{2-\theta}\,,\label{Vphi}\\
V^2(T) &=& \left(1-\frac{2\epsilon}{3q}\right)H^{2(2-\theta)}\,,\label{Vtac}\\
H'a' &=& -\frac{3q}{2}\ell |H|^{\widetilde{\theta}}Ha\,,\label{hj}
\ea
where we have set $\beta_q=1$, $\phi$ is an ordinary scalar field, $T$ is a Dirac-Born-Infeld (DBI) tachyon, $\ell=1$ for ordinary causal fields and $\ell=-1$ for phantoms. Also, $\widetilde{\theta}=\theta\equiv 2(1-q^{-1})$ for the ordinary scalar and $\widetilde{\theta}=2$ for the tachyon, and the parameter
\be 
\epsilon \equiv -\frac{d \ln H}{d \ln a}= -\frac{a}{a'}\frac{H'}{H}\,,\label{epsi}
\ee
defines the time variation of the Hubble radius $R_H\equiv H^{-1}$, where derivatives of the inflaton field are denoted with primes. We will use the symbol $\vp$ to indicate the inflaton field in expressions valid for both the normal scalar and the tachyon. It is convenient to define the new parameter
\be\label{ve}
\ve \equiv \frac{\epsilon}{q|H|^\theta}= \frac{3}{2}\ell\left(\frac{a}{a'}\right)^2\,;
\ee
then one can express the spectral amplitudes as $A_s(\phi)^2\propto H^2/\ve$, $A_s(T)^2\propto H^\theta/\ve$ and get the spectral indices from the evolution equation $\dot{\ve}=2H\ve(\epsilon-\eta)$, which reproduces the 4D one when $\ve=\epsilon$.

The Hamilton-Jacobi equation (\ref{hj}) can be recast as
\be
y(\vp)'a(\vp)'=- y(\vp)a(\vp)\,,\label{hj2}
\ee
where the variable $y(\vp)$ is
\ba 
y(\phi) &=& H(\phi)^{2\ell/3}\,,\qquad\qquad\qquad \theta=0\,,\\
y(\vp) &\equiv& \exp \left[\alpha |H(\vp)|^{-\widetilde{\theta}}\right]\,,\qquad \widetilde{\theta} \neq 0\,,\label{y}
\ea
and the coefficient $\alpha \equiv -2\ell/(3q\widetilde{\theta})$ is $\alpha=-1/3$ and $\alpha=1$ for a RS and GB braneworld without phantoms, respectively.

This definition differs slightly from that of \cite{cal9} in order to get a simpler Hamilton-Jacobi equation. From now on we will set $\widetilde{\theta}=\theta$ for lighter notation. The parameter $\epsilon$ can be written as
\be
\epsilon = -\frac{(\ln y)'^2}{\theta \ln y}\,.
\ee
Let us now consider what transformations are symmetries of Eq. (\ref{hj2}). In general, a symmetry transformation can be written as
\bs\label{map1}\ba
\da(\vp)&=&f_1(\vp)\,,\\
\dy(\vp)&=&f_2(\vp)\,,\label{map1b}
\ea\es
provided that $[\ln f_1(\vp)]'[\ln f_2(\vp)]'=-1$. In Eq. (\ref{map1b}) all the elements of $\dy$, including $\theta$, are evaluated in the dual patch. Since in principle it is not possible to set $\beta=1=\bar{\beta}$ consistently, one should restore the dimensional factors in the previous and following expressions, noting that $[\beta]=E^{(\theta+2)/(\theta-2)}$.

A simple realization of Eq. (\ref{map1}) is
\bs\ba
\da(\vp)&=&y(\vp)^{p(\vp)}\,,\\
\dy(\vp)&=&a(\vp)^{1/p(\vp)}\,.
\ea\es
In order to satisfy the above integrability condition, the function $p(\vp)$ must be either a constant or
\be \label{p}
p(\vp)=p_0\frac{\ln a(\vp)}{\ln y(\vp)}\,,
\ee
where $p_0$ is an arbitrary real constant.

The case of constant $p=p_0$ was considered in \cite{cal9} for $\bar{q}=q$ and $p_0=3/2$, and in \cite{CLLM} for general $p_0$ and energy dependence:
\bs\label{map2}\ba
\da(\vp)&=&y(\vp)^{p_0}\,,\label{map2a}\\
\dy(\vp)&=&a(\vp)^{1/p_0}\,.
\ea\es
The set of equations describing the dual solution can be obtained from Eqs. (\ref{map2a}), (\ref{y}) and (\ref{ve}):
\ba
\da(\vp) &=& \exp \left(-p_0\int^\vp d\vp \frac{a}{a'}\right)\,,\\
|\dH(\vp)| &=& \left[\frac{\dalp p_0}{\ln a(\vp)}\right]^{1/\dth},\label{Hdual}\\
\bar{\ve}(\vp)\,\ve(\vp) &=& \frac{9\ell\bar{\ell}}{4p_0^2}\,.
\ea
The transformation (\ref{map2}) connects the scale factor of the expanding cosmology to that of a dual cosmology when expressed in terms of the scalar field. In the dual model, the scalar field acquires a different time dependence relative to its expanding counterpart. The time variable can be written as an integral over $\vp$, $t=\int^\vp d\vp (\ln a)'/H$; the time variable $\bar{t}$ of the dual solution is then
\be
\bar{t} = \frac{2\ell p_0}{3}\int^\vp \frac{d\vp}{a^{3\bar{\ell}/(2p_0)}} \frac{H'}{H}\,,\label{4Dt}
\ee
in the 4D$\to$4D case,
\be
\bar{t} = \frac{2\ell p_0}{3}\int^\vp d\vp\, (\ln a)^{1/\dth}(\ln H)'\,,
\ee
for the pure braneworld dual of the 4D scenario, while for a general cross duality with $\dth\neq0\neq\theta$, using Eqs. (\ref{y}) and (\ref{hj2}),
\be
\bar{t} = -\frac{p_0}{(\dalp p_0)^{1/\dth}}\int^\vp d\vp \frac{(\ln a)^{1/\dth}}{(\ln a)'}\,.\label{t}
\ee
Concrete computations may show disquieting features when $\dth=\theta \neq 0$. For example, in the Randall-Sundrum scenario a power-law expansion, $a(t)=t^n$, is realized by
\bs \label{rs}\ba
\phi(t) &=&t^{1/2}\,,\qquad V(\phi)=V_0\phi^{-2}\,,\\
a(\phi) &=&\phi^{2n}\,,\qquad H(\phi)=n\phi^{-2}\,,
\ea\es
for a normalized ordinary scalar field, while the dual RS cosmology has 
\bs\label{RSdual}\ba
\da &=& \exp[-p_0\phi^2/(3n)]\,,\\
|\dH| &=& -p_0[3n\ln \phi^2]^{-1}\,,\\
\de&=& -3n(2p_0\phi^2 \ln \phi^2)^{-1}\,.
\ea\es
In the region $\phi<1$ with $p_0>0$ (no phantoms, $\de>0$), the dual scale factor $\da(\phi)$ increases from $\da(1)$ to $\bar{a}(0)$ in a finite time interval; therefore one might regard this as just a portion of a complete solution, were it not for the dual Hubble parameter which actually goes from infinity to zero in the meanwhile. Solutions with $\phi>1$ and $p_0<0$ behaves better, since they are not singular at the origin, and extend up to the infinite future.

Similarly, in the Gauss-Bonnet case we have
\bs\ba
\phi(t) &=& t^{-1/2}\,,\qquad V(\phi)=V_0\phi^6\,,\\
a(\phi) &=& \phi^{-2n}\,,\qquad H(\phi)=n\phi^2\,,
\ea\es
together with the GB dual
\bs\label{GBdual}\ba
\da &=& \exp(p_0 n\phi^2)\,,\\
\dH &=& -(n/p_0)\ln \phi^2\,,\\
\de &=& -(p_0n\phi^2\ln \phi^2)^{-1}\,,
\ea\es
and again the cyclic solution with ordinary matter evolves with $\da<\infty$ for all $t$ and $p_0>0$. On the contrary, in the branch with $p_0<0$ the dual scale factor $\da$ does not collapse to zero at the origin and diverges in the infinite future.

Things do not change when exploring cross dualities. We can try to see what happens, say, for the GB dual of a RS cosmology ($\theta=1$, $\dth=-1$). Starting from Eq. (\ref{rs}), one gets Eq. (\ref{GBdual}) with $p_0 \to -p_0$, modulo an irrelevant positive constant. The image of the function $\phi(\bar{t})$ is either $\{\phi<1\}$ or $\{\phi>1\}$.

\begin{figure}
\includegraphics[width=8.6cm]{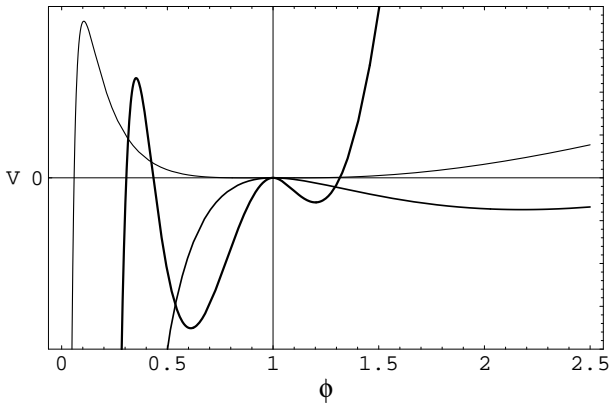}
\caption{\label{fig1}
Gauss-Bonnet potential dual to GB power-law inflation under the mapping (\ref{map2}), for some values of $n$ and $p_0$. The region with $\phi<1$ ($\phi>1$) corresponds to duals with $p_0>0$ ($p_0<0$).}
\end{figure}

Dual potentials can be obtained via the dual of Eq. (\ref{Vphi}) or (\ref{Vtac}). Figure \ref{fig1} shows the potential corresponding to the cosmology Eq. (\ref{GBdual}). Depending on the choice of the parameters $n$ and $p_0$, the function $\bar{V}(\phi)$ has a number of local minima and maxima, can assume negative values and also be unbounded from below.

The properties of cosmological potentials change very interestingly when going from the 4D picture to the braneworld. Take as examples fast-roll inflation with a standard scalar field and negative potentials \cite{lin01,FFKL}. Fast-roll inflation occurs by definition when the kinetic energy of the scalar field is small with respect to the potential energy, $\dot{\phi}^2 \gg V(\phi)$. In this case one obtains a stiff equation of state ($p=\rho$) and a regime described by
\be
a \sim t^{1/3}\,,\qquad \dot{\phi}^2 \sim t^{-2}\,,\qquad \phi \sim \ln t\,,
\ee
from the Klein-Gordon and Friedmann equations. This implies that at early times the kinetic term dominates over any monomial potential energy $V=\phi^m$. In particular, under these conditions the behaviour of the singularity will depend on the kinetic energy independently of the choice of the potential. In a generic patch with $\theta \neq 0$, the fast-roll regime is given by
\be
a \sim t^{1/(3q)}\,,\qquad \dot{\phi}^2 \sim t^{\theta-2}\,,\qquad \phi \sim t^{\theta/2}\,.
\ee
Thus the scalar field evolves quite differently in the RS ($\theta=1$) and GB ($\theta=-1$) case. Near the origin $t\sim 0$ the fast-roll regime is achieved for any $\theta \neq 0$ when $m=2$ and for $\theta>4/(2-m)$ when $m>2$. Therefore the behaviour of the singularity may depend nontrivially on both the contributions of the energy density for suitable (and still simple) potentials on a brane.

Another result in four dimensions is that potentials with a negative global minimum do not lead to an AdS spacetime. According to the Friedmann equation $H^2=\rho$, the energy density cannot assume negative values; therefore at the minimum $V_\text{min}<0$ the scalar field does not oscillate-and-stop but increases its kinetic energy until this dominates over the potential contribution. Then one can describe the instability at the minimum in the fast-roll approximation through the only kinetic tern; the Hubble parameter vanishes and becomes negative [so that $(\dot{\phi}^2/2)^\cdot>0$], and the Universe undergoes a bounce.

In a braneworld scenario this might not be the case. In fact, in the RS brane the Friedmann equation is $H^2=\rho [1+\rho/(2\lambda)]$. If the negative minimum is larger than the brane tension, $|V_\text{min}| \gtrsim \lambda$, then, after an eventual fast-roll transition, the quadratic corrections dominate near the minimum and $H^2 \approx \rho^2$. The scalar field can relax without spoiling the constraints from the equations of motion.

All that we have said can be investigated in greater detail by means of phase portraits in the three-dimensional space $(\phi,\dot{\phi},H)$. Here we will not explore the subject further and limit ourselves to the above qualitative comments, whose aim was to stress that complicated dual potentials cannot be discarded by general classical or semiclassical considerations. Rather, from one side they should be studied case by case; from the other side, one or more local features encountered by the scalar field during its evolution could induce interesting phenomena at the quantum level, for instance triggering premature reheating or a series of quantum tunnelings.

Anyway, it is not yet clear whether the dual solutions constructed so far, especially those with $p_0>0$, describe reasonable (not to mention viable) scenarios. At this point there are two possibilities. The first one is to accept these non-superaccelerating cosmologies and try to explain them by means of some deeper and still missing theoretical ingredient. The second one is to consider their exotic behaviour as a signal that we cannot impose $p_0>0$ (or even $p=$ const.) consistently in pure high-energy braneworlds (at least in the RS and GB cases), while the 4D cosmology can be dual to another 4D cosmology. This is due to the fact that the functions $a(\vp)$ and $y(\vp)$ live in different image real sets. Then a new path to follow is to find some mechanism which ``regularizes'' the dual solutions at the asymptotic past and future. The only degree of freedom we could exploit is given by the parameter $p$, which by this line of reasoning must depend on $\vp$. Therefore we are forced to assume Eq. (\ref{p}), which generates the transformation
\bs\label{map3}\ba
\da(\vp)&=&a(\vp)^{p_0}\,,\label{map3a}\\
\dy(\vp)&=&y(\vp)^{1/p_0}\,.
\ea\es
For $\dth\neq0\neq\theta$ the other dual quantities read
\bs \label{altdu1}\ba
\dH(\vp) &=& \left(\frac{p_0 \dalp}{\alpha}\right)^{1/\dth} H(\vp)^{\theta/\dth}\,,\\
\bar{\ve}(\vp) &=& \frac{\ell\bar{\ell}}{p_0^2}\,\ve(\vp) \quad\Rightarrow\\
\de(\vp) &=& \frac{\theta}{\dth p_0}\,\epsilon(\vp)>0\,,
\ea\es
while
\be \label{altdu2}
\bar{t} = p_0 \left(\frac{\alpha}{p_0\dalp}\right)^{1/\dth}\int^\vp d\vp \frac{(\ln a)'}{H^{\theta/\dth}}\,,%
\ee
so that $\bar{\vp}(t)=\vp(t)$ when $\dth=\theta$. 

In 4D ($\dth=\theta=0$), $\dH(\vp)=H(\vp)^{\ell\bar{\ell}/p_0}$, $\bar{t}=p_0\int^\vp d\vp (\ln a)'/H^{\ell\bar{\ell}/p_0}$ and $\de =\ell\bar{\ell}\epsilon/p_0^2$. The cross duality between the general-relativistic framework ($\theta=0$) and a high-energy braneworld ($\dth\neq 0$) is, after a time redefinition,
\bs\ba
\dH(\vp) &=& [\ln H(\vp)]^{-1/\dth}\,,\\
\de(\vp) &=& -\epsilon(\vp)[p_0\dth\ln H(\vp)]^{-1}\,,\\
\bar{t}  &=& p_0 \int^\vp d\vp\, (\ln H)^{1/\dth}(\ln a)'\,.
\ea\es
Clearly, the effect of Eqs. (\ref{map3}), (\ref{altdu1}) and (\ref{altdu2}) results in a rescaling of time when $\dth=\theta$, as one can verify by making the substitution
\be
p_0 \to p_0 \frac{\ln \phi^2}{\phi^2}\,,
\ee
in the RS and GB power-law duals, Eqs. (\ref{RSdual}) and (\ref{GBdual}). In this case (which includes tachyon-tachyon dualities) duals without phantoms are achieved as long as $p_0>0$. 

When $\dth\neq\theta$, this transformation relates the dynamics of different braneworld scenarios. According to the cross duality between RS and GB standard inflation, the dual solution does not superaccelerate if, and only if, $p_0<0$. The power-law case is trivial since the dual GB solution is 
\be
\da = \phi^{-2\bar{n}}\,,\qquad \dH = \phi^2\,,\qquad \de = \bar{n}^{-1}\,,
\ee
where $\bar{n}\equiv -np_0$ and $\phi(t) \propto t^{-1/2}$. 

In the power-law case the mapping (\ref{map3}) can be realized also by
\bs\label{map4}\ba
\da(\vp) &=& [\ln y(\vp)]^s\,,\\
\dy(\vp) &=& \exp\left(\frac{1}{s\theta}\int^\vp d\vp \frac{H}{H'}\right)\,,
\ea\es
where $s$ is a real constant, giving a power-law dual $\da = t^{|s|}$. Note the domain range of the dual scale factor. The dual parameter $\de$ is
\be
\de = \frac{\alpha}{\dalp\dth\theta s^2}\frac{|\dH|^{\dth}}{|H|^\theta}\frac{1}{\epsilon}\,,
\ee
which shows how in general the mapping (\ref{map4}) is not equivalent to Eq. (\ref{map3}). This can be seen also by considering the action of the former in four dimensions, where the dual Hubble parameter reads
\be
\dH = \exp\left[-\frac{3\bar{\ell}}{2 s}\int^\phi d\phi\frac{\ln H}{(\ln H)'}\right]\,.
\ee
The power-law solution is $a=\exp(n\phi)$, $H=\exp(-\phi)$ and its dual is $\da =\phi^s$, $\dH =\exp (-\phi^2/s)$ and $\de=2\phi^2/s^2$, with potential $\bar{V}=(1-\de/3)\exp(-s\de)$. If $s<0$, there is an instability as $\phi \to \infty$, while for positive $s$ the potential has a local minimum at $\de_*=3+1/s$ [$V''(\de_*) \propto s$] and vanishes at large $\phi$. 

One can devise other transformations of the Hamilton-Jacobi equation than Eqs. (\ref{map2}), (\ref{map3}) and (\ref{map4}). The last example we give is the following:
\bs\label{map5}\ba
\da(\vp) &=& \exp\left(-\frac{1}{r}\int^\vp \frac{d\vp}{a'}\right)\,,\\
\dy(\vp) &=& \exp [r a(\vp)]\,,
\ea\es
where $r$ is a real constant. For $\dth \neq 0\neq \theta$, the basic equations are
\be
|\dH|=\left(\frac{\dalp}{ra}\right)^{1/\dth},\qquad \de= -\frac{r}{\dth}\frac{a'^2}{a}\,,\qquad \dot{\bar{\vp}}=-ra'\left(\frac{\dalp}{ra}\right)^{1/\dth}.
\ee
The RS$\to$RS dual ($r<0$) has $\da \sim \exp t^{1-n}$, $\dH\sim t^{-n}$ and $\de \sim t^{n-1}$. The RS$\to$GB dual ($r>0$) has $\da \sim \exp t^{(1-n)/(1-2n)}$, $\dH\sim t^{n/(1-2n)}$ and $\de \sim t^{(1-n)/(1-2n)}$. The GB$\to$GB dual ($r>0$) has $\da \sim \exp t^{(1+n)/(1+2n)}$, $\dH\sim t^{-n/(1+2n)}$ and $\de \sim t^{-(1+n)/(1+2n)}$. 

In the limit $n \to \infty$, the GB dual of both RS and GB cosmology is $\da \sim \exp \sqrt{t}$, that is the Randall-Sundrum self-dual solution with respect to Eq. (\ref{map2}).

We conclude with an interesting remark. The above dualities connect not only different braneworlds with the same type of scalar field but also patches with different scalars. If one wishes to construct cosmologies with a DBI tachyon, it is sufficient to start from a generic scenario $(q,\widetilde{\theta},\vp)$ and hit the dual $(\bar{q},\dth=2,T)$ via either Eq. (\ref{map2}), (\ref{map3}), (\ref{map4}) or (\ref{map5}). In particular, with Eq. (\ref{map3}) 
\ba
\dH(T) &=& H(\vp \to T)^{\theta/2}\,,\qquad\qquad \theta\neq 0\,,\\
\dH(T) &=& [\ln H(\vp \to T)]^{-1/2}\,,\qquad \theta= 0\,,
\ea
in agreement, e.g., with previous results on power-law standard and tachyon inflation (see \cite{cal3} and references therein).

In this Letter we have established several relations linking cosmological models living in a number of geometrical and dynamical realizations, including the 4D, RS and GB pictures as well as tachyonic scenarios. Hopefully, the cross dualities $(\theta,\vp)  \leftrightarrow (\dth,\bar{\vp})$ we have found will help in the understanding of high-energy phantom and bouncing models of the early Universe.


\end{document}